\documentclass[pra, twocolumn, showpacs, superscriptaddress]{revtex4-1}

\usepackage{graphicx,amsmath,amssymb}

\begin{document}

\title{Radio frequency spectrum of fermions near a narrow Feshbach resonance}
\author{Junjun Xu}
\affiliation{Department of Physics, University of Science and Technology Beijing, Beijing 100083, China}
\affiliation{Laboratory of Atomic and Solid State Physics, Cornell University, Ithaca, New York 14853, USA}
\author{Qiang Gu}
\affiliation{Department of Physics, University of Science and Technology Beijing, Beijing 100083, China}
\author{Erich J. Mueller}
\affiliation{Laboratory of Atomic and Solid State Physics, Cornell University, Ithaca, New York 14853, USA}
\date{\today}

\begin{abstract}
We calculate the radio frequency (RF) spectrum of fermionic atoms near a narrow Feshbach resonance,
explaining observations made in ultracold samples of $^6\rm{Li}$ [E. L. Hazlett {\it et al.}, Phys. Rev. Lett.  {\bf 108}, 
045304 (2012)]. We use a two channel resonance model to show that the RF spectrum contains
two peaks. In the wide-resonance limit, nearly all spectral weight lies in one of these peaks, and typically
the second peak is very broad. We find strong temperature dependence, which can be traced to the
energy dependence of the two-particle scattering. In addition to microscopic calculations, we use sum
rule arguments to find generic features of the spectrum which are model independent.
\end{abstract}
 
\maketitle

\section{Introduction}
Magnetic field induced scattering resonances give cold atom experiments the ability to tune the
strength of inter-atomic interactions \cite{Ketterle, Bloch, Chin}. For example, for fields in the range
$824-844\rm{G}$, the scattering length in a gas of fermionic $^6\rm{Li}$ changes by several 
orders of magnitude. Studies of superfluidity at these fields have revolutionized our understanding
of the connections between BCS-pairing of fermions and Bose-Einstein condensation of composite
bosons. Recent attention has turned to ``narrow" resonances, where the characteristic field over which
the scattering length changes is $\sim0.1\rm{G}$ \cite{Strecker, Schwenk, Petrov, Ho, Ohara, Jensen}. 
As will be described below, the scattering properties near narrow resonances are more complicated,
featuring energy dependences which are not captured by the scattering length. Here we study how
this energy dependence is manifest in radio frequency (RF) spectra.

When the de Broglie wavelength of an atomic gas is large compared to the range of interactions,
one is in the cold-collision limit, and all scattering properties are encoded in the s-wave scattering
amplitude ${\rm Re}(f_0(k))=-a_s-r_{\rm{eff}}a_s^2k^2+O(k^3)$. The scattering cross section between
particles with relative momentum $k$ is proportional to $|f_0(k)|^2$. Low energy scattering is typically
characterized by the s-wave scattering length $a_s=-f_0(k=0)$. The scattering length is a function
of magnetic field, diverging at the Feshbach resonance field $B_0$, with the functional form
\begin{eqnarray}
a_s(B)=a_{\rm{bg}}(1-\frac{\Delta B}{B-B_0}).
\end{eqnarray}
Here, $\Delta B$ is the width of the resonance, and $a_{\rm{bg}}$ is the background scattering length,
describing the scattering far from resonance. These resonances are generically associated with
a crossing between a ``closed channel" molecular state and the open-channel continuum. The 
characteristic scale over which $f_0$ changes is given by 
$r_{\rm{eff}}=\hbar^2/m\mu a_{\rm{bg}}\Delta B$ \cite{Petrov}, where $\mu\approx\mu_B$ is the
difference in the magnetic moments of the two channels. If $r_{\rm{eff}}k\gg1$ for a typical collision,
then the scattering length is insufficient to describe the physics.

Recently, Ho {\it et al.} have pointed out that for a narrow resonance, because of the energy dependence
of the phase shift, the interaction energy is highly asymmetric and strong interactions persist even for
$B-B_0\gg\Delta B$ on the BCS side \cite{Ho}. This observation is consistent with the studies of Jensen
{\it et al.} at the impact of the effective range on the thermodynamics of the BCS-BEC crossover
\cite{Jensen}, and few-body studies by Petrov \cite{Petrov}. Schwenk and Pethick used related arguments
to constrain the equation of state of nuclear matter \cite{Schwenk}.

Following these theoretical developments, O'Hara's experimental group has  studied a narrow resonance
in $^6{\rm Li}$, finding that the interaction energy and three-body recombination rate are both strongly
energy dependent \cite{Ohara}. This energy dependence can lead to novel many-body physics, such as
breached-pair superfluidity \cite{Liu}.

A similar experiment with ${\rm Li}-{\rm K}$ mixtures has been performed by Kohstall {\it et al.}
\cite{Kohstall}. They too study the RF spectrum near a narrow resonance, with extra complications due to
the disparate masses and densities of the two species. Here we restrict our discussion to the simpler
homonuclear problem. Qualitatively, their observations are very similar to O'Hara's. In this paper, we will
calculate the RF spectrum of $^6{\rm Li}$ atoms near the
narrow resonance around 543G. As in the experiment of Hazlett {\it et al.}, we consider the system
initially in the lowest and third lowest hyperfine state (defined as 1 and 3). The Feshbach resonance
does not couple these atoms, and the system is readily modeled as non-interacting. RF waves will
induce a transition between 3 and the second lowest hyperfine state (defined as 2). The shape of
the absorption line will be modified by the interactions between atoms in state 1 and 2. Consequently
the absorption spectrum will have strong dependence on the magnetic field. One hopes to use details
of the RF lineshape to learn about the underlying physics \cite{Mueller1, Mueller2, Stewart, He, Pieri1,
Pieri2, Haussmann, Greiner, Chin2, Schunck, Cheuk}. This program is analogous to how the tunneling
spectra in superconductors can reveal features of the phonon pairing potential \cite{McMillan}.

To calculate the lineshape we sum an infinite set of diagrams, restricting ourselves to intermediate states
without particle-hole excitations. Similarly we do not include the inelastic decay of the excited Feshbach
molecules. These latter processes should slightly broaden the spectrum. This approach yields relatively
simple results, and obeys all of the appropriate sum rules. Including more complicated intermediate states
will quantitatively change the detailed lineshape, leaving gross features (such as its first few moments)
unchanged.

Through out this paper we restrict ourselves to a uniform gas whose density corresponds to the
average density of the experimental harmonically trapped system. A more sophisticated treatment
would include inhomogeneous broadening from the trap \cite{Mueller}.

Our paper is organized as follows: We first introduce the two channel resonance model which
describes the system. Then we give a simple sum rule argument to extract generic features of
the RF lineshape as one changes the resonance width. Next we calculate the RF spectrum
from a $T=0$ variational ansatz. Next we generalize our calculation to finite temperature
using Matsubara Green's function techniques. Finally, we compare our results with experiments. 

\section{model}
To describe the 3-component fermions near a narrow Feshbach resonance, we use the following
two channel resonance model \cite{Timmermans}
\begin{eqnarray}
H &&= \sum_{k, \sigma} \epsilon_{k, \sigma} a_{k, \sigma}^\dagger a_{k, \sigma} + \sum_{k} 
\left( \epsilon_{k, b} + \delta \right) b_k^\dagger b_k \nonumber\\ 
&&+\frac{\lambda}{\sqrt{\Omega}}\sum_{p, q} \left( b_{p+q}^\dagger a_{p, 1} a_{q, 2} + h.c. \right),
\label{eq:twochannelmodel}
\end{eqnarray}
where the first term in the Hamiltonian corresponds to the energy of isolated atoms: $a_{k,\sigma}$
annihilates an atom with momentum $k$ and spin $\sigma=1,2,3$, whose energy is 
$\epsilon_{k, \sigma}=\hbar^2k^2/2m-\mu_\sigma$, where $\mu_{\sigma}$ is the chemical 
potential. The second term corresponds to the energy of isolated molecules,
$\epsilon_{k,b}=\hbar^2k^2/4m-\mu_1-\mu_2$ and $\delta=2\mu_B(B-B_\infty)$ is the detuning
between the open and closed channel , where $2\mu_B$ is the magnetic moment difference between
open and closed channel in $^6$Li with $\mu_B$ the Bohr magneton. The last term in the Hamiltonian 
$\lambda\Lambda/\sqrt{\Omega}=\lambda\sum_{p, q} \left( b_{p+q}^\dagger a_{p, 1} a_{q, 2} + h.c. \right)/\sqrt{\Omega}$
parameterizes the coupling between open and closed channels via a single coefficient
$\lambda$. $\Omega$ is the volume of the system. To second order in $\lambda$, the two-body T-matrix
describing scattering between states 1 and 2 is $T^{2B}(k)=\frac{\lambda^2}{\Omega(2\epsilon_k-\delta)}$,
where $\epsilon_k=\hbar^2k^2/2m$ is the energy of one particle before scattering. Thus the s-wave
scattering length of the system is
\begin{eqnarray}
a_s=\frac{m\Omega}{4\pi\hbar^2} T^{2B}(0)=-\frac{m\lambda^2}{4\pi\hbar^2\delta},
\label{eq:as}
\end{eqnarray}
which can be compared with the empirical magnetic field dependence
$a_s=a_{\rm{bg}}(1-\frac{\Delta B}{B-B_\infty})\approx -a_{\rm{bg}}\frac{\Delta B}{B-B_\infty}$. Hence
$\lambda$ is related to the experimental observables via
\begin{eqnarray}
\lambda=\sqrt{\frac{8\pi\hbar^2a_{\rm{bg}}\Delta B\mu_B}{m}}.
\label{eq:lambda}
\end{eqnarray}

As introduced in section I, the spin states $\sigma=1,2,3$ model the three lowest energy hyperfine states
of $^6{\rm Li}$ near the narrow Feshbach resonance at $B=543{\rm G}$. Inserting known experimental
parameters for $^6\rm{Li}$ with $a_{\rm{bg}}\approx62a_0$ and $a_0$ the Bohr radius, we find
$\lambda\approx2.9\times10^{-39}\rm{J\sqrt{m^3}}$. The effective range of the model is
$r_{\rm{eff}}=4\pi\hbar^4/m^2\lambda^2\approx 3.5\times 10^4a_0$. For a uniform gas of $^6\rm{Li}$
gas with density $n=10^{13}\rm{cm}^{-3}$, the Fermi wave vector
$k_F=(6\pi^2n)^{1/3}\approx8.4\times10^6\rm{m}^{-1}$. In this case, we have $r_{\rm{eff}}k_F\approx15.4$,
corresponding to a narrow resonance.

At time $t=0$, we imagine the system is prepared with an equal number of particles in states 1 and 3,
$N_1=N_3=N$, and no particles in state 2, $N_2=0$. Within our model, interactions vanish for this
initial state. To investigate the narrow resonance between states 1 and 2, we introduce a radio frequency
probe which drives atoms from state 3 into state 2. This probe can be modeled by a perturbation
\begin{eqnarray}
V=\sum_k\left(a_{k,2}^\dagger a_{k,3}e^{-i\omega^\prime t}+a_{k,3}^\dagger a_{k,2}e^{i\omega^\prime t}\right),
\label{eq:pert}
\end{eqnarray}
where $w^\prime=w-(\mu_2-\mu_3)$. The physical radio waves have frequency $\nu=\nu_0+w/h$ where
$\nu_0$ is the free-space resonance frequency for the transition from state 3 to 2. 
For simplicity, we use units where $\hbar=k_B=1$ and denote $\mu_1=\mu_3=\mu$, $\mu_2=0$.
Thus in our model $w^\prime=w+\mu$.

\subsection{Sum rules}
At zero temperature, the ground state of our system (in the absence of the probe) is a  Fermi sea 
of equal numbers of $1$ and $3$ particles 
$|GS\rangle=|F\rangle=\Pi_{k<k_F}a_{k,1}^\dagger a_{k,3}^\dagger|0\rangle$. 
The probe in Eq. (\ref{eq:pert}) generates transitions from state 3 to state 2 at a rate
\begin{eqnarray}
I(w)&&=2\pi\sum_f \left| \langle GS|V|f \rangle \right|^2 \delta\left(w+\mu-E_f+E_0\right)
\nonumber\\
&&\propto {\rm Im}\langle GS | V \frac{1}{w-\bar{H}} V^\dagger | GS \rangle,
\label{eq:specformu}
\end{eqnarray}
where $\bar{H}=H-E_0-\mu$. The energy of the ground state is $E_0$, and the sum is over all final 
states $|f\rangle$ with energy $E_f$. In this subsection, we calculate moments of $I(w)$. Our results
will be exact. We will then use these moments to describe qualitative features of the spectrum.

First, the total spectral weight is simply given by the number of atoms initially in state 3,
\begin{eqnarray}
S_0=\int \frac{dw}{2\pi}I(w)&&=\sum_f|\langle GS|V|f \rangle|^2\nonumber\\
&&=\langle GS|VV^\dagger |GS \rangle\nonumber\\
&&=N.
\label{eq:sr0}
\end{eqnarray}
Second, the first moment vanishes
\begin{eqnarray}
S_1=\int \frac{dw}{2\pi}wI(w)=\langle GS|V\bar{H}V^\dagger |GS \rangle=0,
\label{eq:sr1}
\end{eqnarray}
implying  that the spectrum should extend over both the negative and positive RF frequencies with a
centroid at $w=0$. Third, the second moment is
\begin{eqnarray}
S_2=\int \frac{dw}{2\pi}w^2I(w)&&=\langle GS|V\bar{H}\bar{H}V^\dagger |GS \rangle\nonumber\\
&&=-\langle GS|[V,\bar{H}]^2|GS \rangle \nonumber\\
&&=\frac{\lambda^2}{\Omega}N^2.
\label{eq:sr2}
\end{eqnarray}
Finally, detuning dependence is encoded in the third order sum rule
\begin{eqnarray}
S_3=\int \frac{dw}{2\pi}w^3I(w)&&=\frac{\lambda^2}{\Omega}\langle GS|V\Lambda\bar{H}_0
\Lambda V|GS \rangle \nonumber\\
&&=\frac{\lambda^2}{\Omega}N^2\delta_{\rm{shift}},
\label{eq:sr3}
\end{eqnarray}
where $\bar{H}_0=\bar{H}-\lambda\Lambda/\sqrt{\Omega}$, $\delta_{\rm{shift}}=\delta-3E_F/5$
with $E_F=k_F^2/2m$ the Fermi energy and $k_F=(6\pi^2N/\Omega)^{1/3}$.

To get a qualitative picture of the spectrum, we imagine a bimodal distribution made up from two
$\delta$-function peaks,
\begin{eqnarray}
I(w)\approx A_+\delta(w-w_+)+A_-\delta(w-w_-).
\label{eq:rfmodel}
\end{eqnarray}
Note: this ansatz does not capture the fact that the peaks may be quite broad and asymmetric. Further, the
frequencies $w_\pm$ should be interpreted as the centroid of the spectral line, rather than the location of 
maximum intensity.

From the sum rules Eq. (\ref{eq:sr0})-(\ref{eq:sr3}) we find
\begin{eqnarray}
A_\pm=\pi N\left(1\mp\frac{\delta_{\rm{shift}}}{\sqrt{\delta_{\rm{shift}}^2+4N\lambda^2/\Omega}}\right),\\
w_\pm=\frac{1}{2}\left(\delta_{\rm{shift}}\pm\sqrt{\delta_{\rm{shift}}^2+4N\lambda^2/\Omega}\right).
\label{eq:sumrules}
\end{eqnarray}

\begin{figure}[th]
  \includegraphics[width=0.44\textwidth]{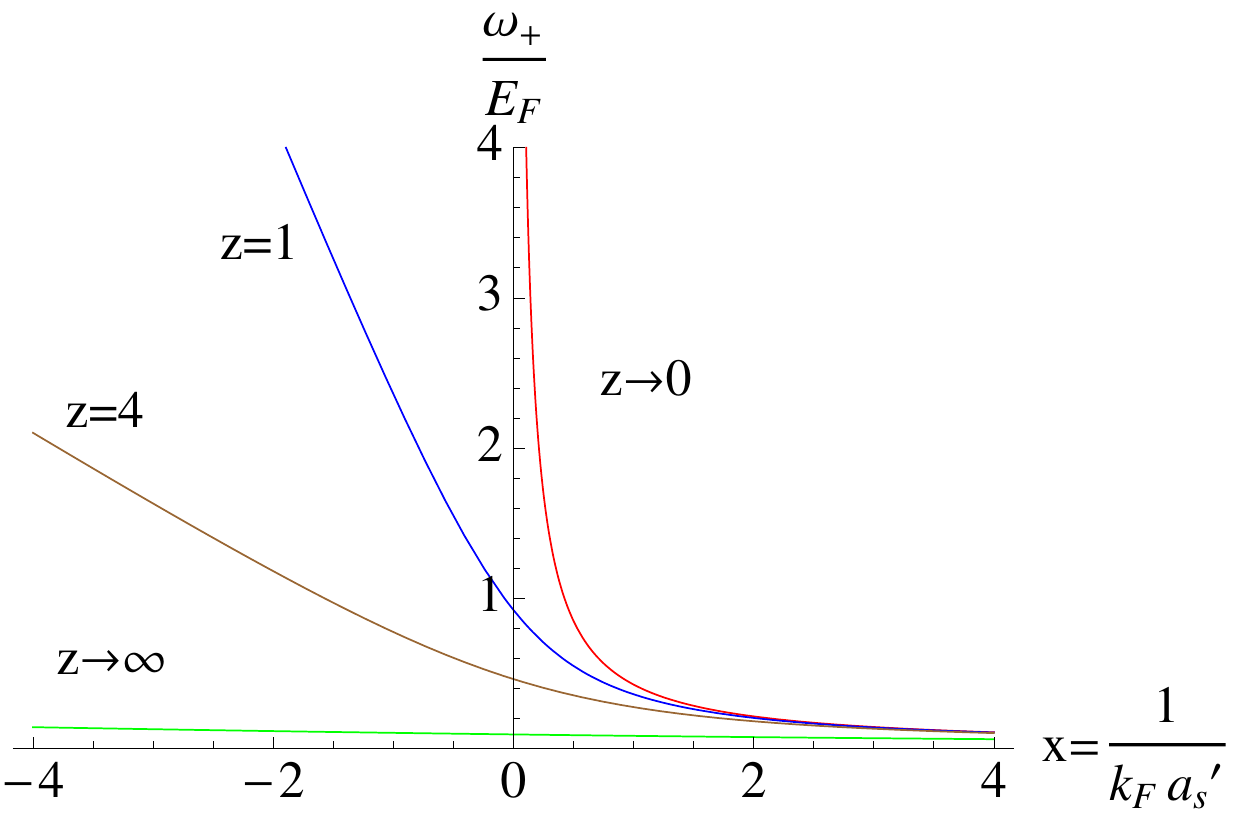}
  \caption{Illustration of the evolution of the RF peaks from wide to narrow resonance. Vertical axis
  shows the frequency of one of the two peaks in the RF spectrum, as estimated by our sum rule
  argument. Different lines are labeled by their value of $z=k_Fr_{\rm{eff}}$. A wide resonance
  corresponds to $z\to 0$.}
  \label{fig:sumrules}
\end{figure}

At $\delta_{\rm{shift}}=0$, the two peaks have equal weight, and it is natural to define an effective
scattering length $a_s^\prime=-m\lambda^2/4\pi\delta_{\rm{shift}}$. In the limit where the Fermi
energy is small compared to the detuning this corresponds to the standard definition in Eq. (\ref{eq:as}).
As will be more precisely described below, for a wide resonance, one almost always has
$E_F\ll\delta$, so $a_s^\prime\approx a_s$.

To illustrate the structure of Eq. (\ref{eq:sumrules}), we rewrite it in terms of the dimensionless
variables $x=1/k_Fa_s^\prime$ and $z=k_Fr_{\rm eff}$,
\begin{eqnarray}
\frac{w_\pm}{E_F}=g_\pm\left(x, z\right),
\end{eqnarray}
where $g_\pm(x, z)=-(x\pm\sqrt{x^2+8z/3\pi})/z$. The variable $x$ is a measure of the interaction
strength, while $z$ is a measure of the resonance width. Fig. \ref{fig:sumrules} shows $w$ as a function
of $x$ for several values of $z$. We only include the positive frequency $w_+$ in Fig. \ref{fig:sumrules}.
The equivalent picture for $w_-$ is generated by noting that $g_-(x,z)=-g_+(-x,z)$. As $x\to 0$, the peak
moves to $w_+=\sqrt{8/(3\pi z)}E_F$. In the wide resonance limit, $z\to0$, the frequency shift
diverges at $x=0$. Additionally, for $x\gg\sqrt{z}$, the coefficients simplify $A_+\to1$ and $A_-\to 0$,
and there is effectively only a single peak, with $w_+\approx4\pi na_s/m$. On the other hand, in the
narrow resonance limit, $z\gg 1$, the peaks have nearly equal weight and disperse slowly as a function
of the scattering length.  The curves in Fig.~\ref{fig:sumrules}  become flatter as the resonance width decreases. 

In summary, the sum rules suggest the following:
\begin{itemize}
\item[(1)] In the limit of a wide resonance, the spectrum is dominated by a single peak whose mean frequency
$w\approx 4\pi na_s/m$, and $w\to\infty$ as $a_s\to\infty$.
\item[(2)] For a finite width resonance this divergence is cut off.
\item[(3)] Generically, the spectrum will be bimodal near resonance.
\item[(4)] The location of the resonance, defined by when equal spectral weight lies in each peak,
is shifted from its free-space value.
\end{itemize}

The divergence in (1) is a manifestation of the similar divergence seen in sum rule calculations of the
mean line-shift in the RF-absorption from a superfluid initial state to a noninteracting final state
\cite{Schneider}. It should be interpreted as a divergence of the first moment of the spectral line, rather
than the location of the peak.

Beyond these generalities, the sum rule arguments do not tell us about the detailed lineshapes.
In the following subsections we present more sophisticated arguments to access these details. We will
find that near resonance the peaks become quite broad, with spectral width growing as the temperature
increases. 

\subsection{Zero temperature}
In the following subsections, we give a quantitative description of the RF spectrum. First we consider
zero temperature, approximating the sum in Eq. (\ref{eq:specformu}) by projecting $\bar{H}$ into a
restricted subspace. In subsection II. C, we show that this projection is equivalent to summing a certain
set of Feynman diagrams.

We consider intermediate states of the form
\begin{eqnarray}
|n\rangle=\left\{
\begin{array}{ll}
|q\rangle=a_{q,2}^\dagger a_{q,3}|F\rangle &(q<k_F) \\
|p,q\rangle=b_{p+q}^\dagger a_{p, 1} a_{q, 3}|F\rangle &(p,q<k_F)
\end{array}\right.,
\end{eqnarray}
where $|F\rangle$ is the filled Fermi sea of atoms in states 1 and 3. The state $|q\rangle$ represents
the situation where the atom in spin-state 3 with momentum $q$, has been transferred into spin-state 2.
This atom can bind with an atom in spin-state 1 with momentum $p$, forming a molecule with momentum
$p+q$, described by state $|p,q\rangle$. We neglect possible intermediate states where this molecule 
then breaks up into a pair of atoms with momentum $p^{\prime}$ and $q^{\prime}$. These latter states
look similar to $|q\rangle$, but have extra particle-hole excitations. In the limit $\lambda\to0$, such
processes are suppressed relative to the terms we keep. Our approximation satisfies the sum rules in
subsection II. A. In this truncated space, the coupling interaction
$\Lambda=\sum_{p, q} \left( b_{p+q}^\dagger a_{p, 1} a_{q, 2} + h.c. \right)$
relates these two states $\sum_p|p,q\rangle=\Lambda|q\rangle$ and $|q\rangle=\Lambda|p,q\rangle$. 
The relevant matrix elements of $\bar{H}$ are
\begin{eqnarray}
&&\langle q|\bar{H}|q\rangle=\epsilon_{q,2}-\epsilon_{q,3}-\mu=0,\\
&&\langle p,q|\bar{H}|p,q\rangle=\delta+\epsilon_{p+q, b}-\epsilon_{p,1}-\epsilon_{q,3}-\mu,\\
&&\langle q|\bar{H}|p,q\rangle=\lambda/\sqrt{\Omega}.
\end{eqnarray}
All other matrix elements vanish. Eq. (\ref{eq:specformu}) can then be cast as a readily summable series
in $\lambda$,
\begin{eqnarray}
I(w)&&\propto {\rm Im}\sum_{nn^\prime}\langle GS | V|n\rangle\langle n| \frac{1}{w-\bar{H}} 
|n^\prime\rangle\langle n^\prime|V^\dagger | GS \rangle\nonumber\\
&&={\rm Im}\sum_{q<k_F}\langle q|\frac{1}{w-\bar{H_0}}\sum_{m=0}^\infty(\frac{\lambda\Lambda}{\sqrt{\Omega}
(w-\bar{H_0})})^{2m}|q\rangle\nonumber\\
&&={\rm Im}\sum_{q<k_F} \frac{1}{w+\mu+\epsilon_{q,3}-\epsilon_{q,2}-\lambda^2\theta(q,w)},
\label{eq:zerotemp}
\end{eqnarray}
where
\begin{eqnarray}
\theta(q,w)=\frac{1}{\Omega}\sum_{p<k_F}\frac{1}{w+\mu+\epsilon_{q,3}+\epsilon_{p,1}-\epsilon_{p+q,b}-\delta}.
\end{eqnarray}

When $\lambda\to0$, $I(w)\to\delta(w)$, corresponding to the
response of free atoms. In section II. D. we numerically calculate the sums and explore the
resulting spectra.

\begin{figure}[tbp]
  \includegraphics[width=0.2\textwidth]{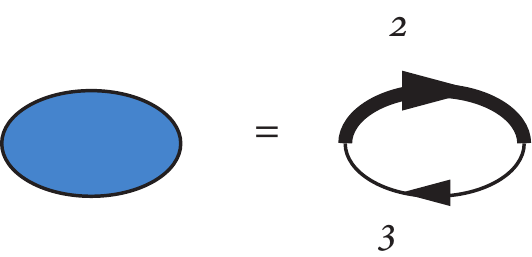}
  \caption{Graphical representation of Eq. (\ref{eq:1a}), corresponding to the exact result for the
  Hamiltonian in Eq. (\ref{eq:twochannelmodel}), that the polarization $R(k, iw_n)$ is the product
  of two Green's functions. Thick lines are dressed propagators while thin ones are bare propagators.}
  \label{fig:grapha}
\end{figure}

\begin{figure}[tbp]
  \includegraphics[width=0.4\textwidth]{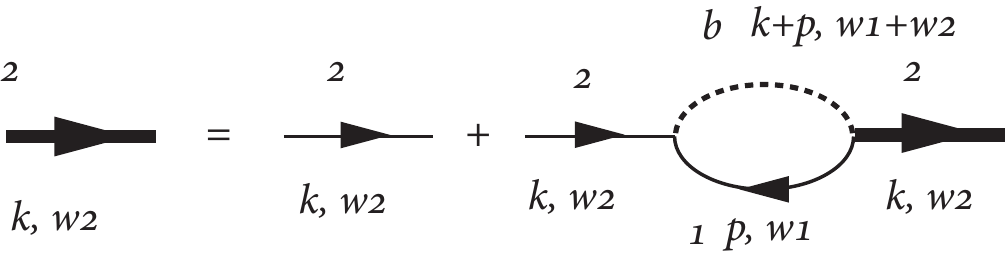}
  \caption{Diagrammatic approximation used in Eq. (\ref{eq:approx}).  
  Solid lines are fermions (particles 1, 2) and dashed lines are molecular states. 
  Thick lines are dressed propagators while thin ones are bare propagators.}
  \label{fig:graphb}
\end{figure}

\begin{figure*}[t]
  \includegraphics[width=1\textwidth]{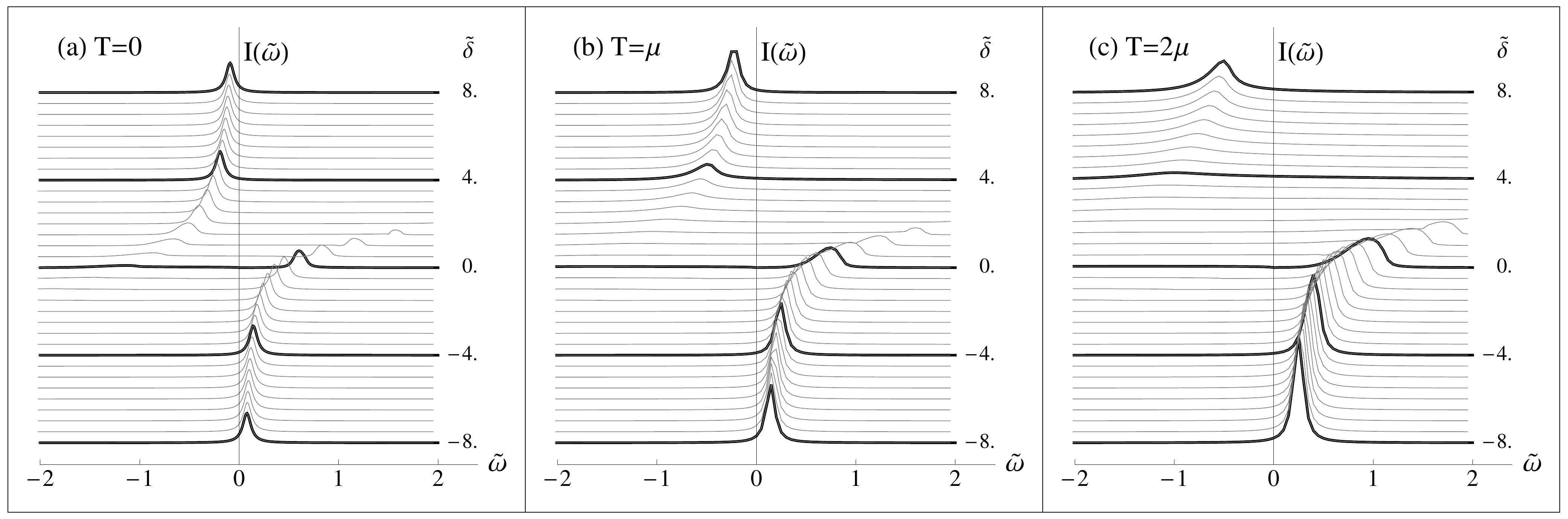}
  \caption{Profiles of the RF spectrum as a function of reduced magnetic detuning $\tilde{\delta}=\delta/\mu$
  and RF frequency $\tilde{w}=w/\mu$ at temperature $T=0$, $\mu$, and $2\mu$.}
  \label{fig:theory}
\end{figure*}

\begin{figure}[t]
  \includegraphics[width=.47\textwidth]{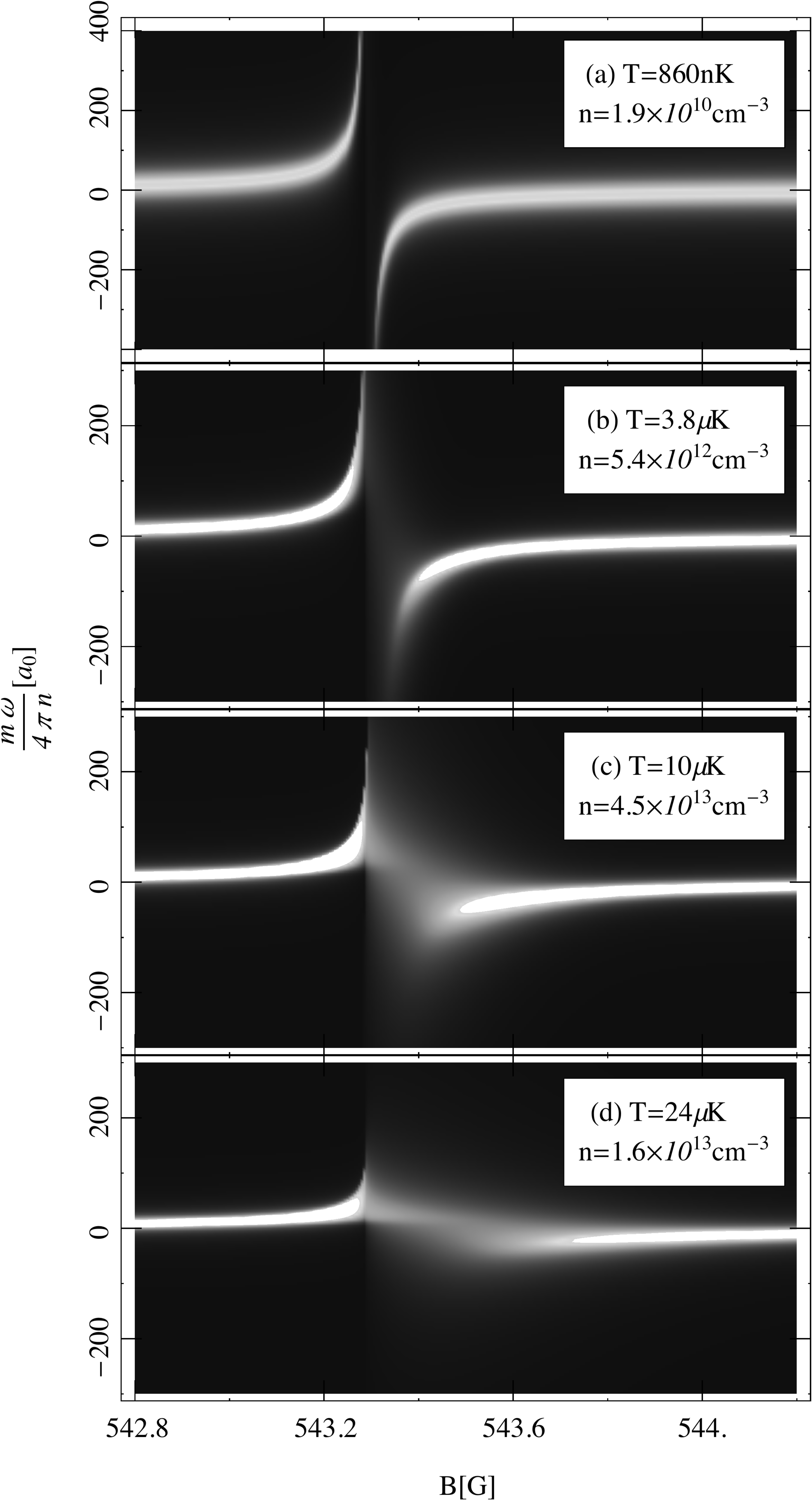}
  \caption{RF spectrum as function of magnetic field B and RF frequency $w$ for different temperatures
  $T=0.86\mu\rm{K}$, $3.8\mu\rm{K}$, $10\mu\rm{K}$, and $24\mu\rm{K}$ and densities 
  $n=1.9\times 10^{10}{\rm cm}^{-3}$, $n=5.4\times 10^{12}{\rm cm}^{-3}$, 
  $n=4.5\times 10^{13}{\rm cm}^{-3}$ and $n=1.6\times 10^{13}{\rm cm}^{-3}$. 
  Brighter color corresponding to higher intensity. Spectra and fields are given in physical units.}
  \label{fig:experiment}
\end{figure}

\subsection{Finite temperature}
In this subsection we generalize our $T=0$ calculation to finite temperature. In particular, the RF spectrum
is given by
\begin{eqnarray}
I(w^\prime)&&\propto {\rm Im}\frac{1}{V}\sum_kR(k, w^\prime+i0)\nonumber\\
&&={\rm Im}\int R(k,\tau)e^{i(w^\prime+i0)\tau}d\tau,
\end{eqnarray}
where
\begin{eqnarray}
R(k,\tau)&&=-\langle T_\tau \psi_3^\dagger(k, \tau)\psi_2(k, \tau)\psi_2^\dagger(k, 0)\psi_3(k, 0) \rangle
\nonumber\\
&&=\langle T_\tau\psi_3(k,0)\psi_3^\dagger(k,\tau)\rangle
\langle T_\tau\psi_2(k,\tau)\psi_2^\dagger(k,0)\rangle
\nonumber\\
&&=G_3(k,-\tau)G_2(k,\tau)
\end{eqnarray}
is the imaginary time retarded Green's function with $\tau=it$ and $\tau>0$. $G_3(k,-\tau)$ and
$G_2(k,\tau)$ are single particle Green's functions for the atoms in states 3 and 2. The brackets
represent thermal expectation values in the absence of the RF coupling, but in the presence of
interactions: $\langle A\rangle={\rm Tr}(Ae^{\beta H})$ with $\beta=1/T$. 
It is more convenient to use the Matsubara representation,
\begin{eqnarray}
R(k,iw_n)&&=\int R(k,\tau)e^{iw_n\tau}d\tau\nonumber\\
&&=\sum_{w_2}\frac{1}{\beta}G_3(k, i(w_2-w_n))G_2(k,iw_2).
\label{eq:1a}
\end{eqnarray}
Here $w_n=2\pi n/\beta$ and $w_\sigma=(2\pi+1)m/\beta$ with $m, n\in{\mathbb N}$. The relationship
is shown diagrammatically in Fig. \ref{fig:grapha}. Since the Hamiltonian contains no
interactions involving particles in state 3, $G_3=1/(iw_3-\epsilon_{k,3})$ is a bare propagator. Using
the standard techniques of many-body perturbation theory, $G_2$ can be expressed as an infinite
sum of diagrams. The natural extension of the approximation in subsection II. B. involves truncating
this sum to only include terms without particle-hole pairs. The resulting series is expressed as a
Dyson sum in Fig. \ref{fig:graphb}. It corresponds to writing the propagator as a geometric series

\begin{eqnarray}
G_2(k, iw_2)=\frac{1}{G_2^0(k,iw_2)^{-1}-\lambda^2\Sigma(k,iw_2)},
\label{eq:approx}
\end{eqnarray}
where $G^0_{\sigma=1,2,3}(k,iw_\sigma)=1/(iw_\sigma-\epsilon_{k,\sigma})$ is the
bare atomic propagator. $\Sigma(k,iw_2)$ is the self energy of particle 2, which we take to be
\begin{eqnarray}
\Sigma(k,iw_2)&&=\frac{1}{V}\sum_{p, w_1}G_b^0(k+p, i(w_1+w_2))G_1^0(p, iw_1)\nonumber\\
&&=\int d^3p\frac{f(\epsilon_{p, 1})}{iw_2+\epsilon_{p, 1}-\epsilon_{k+p, b}-\delta},
\end{eqnarray}
where $G_b^0(k,iw_n)=1/(iw_n-\epsilon_{k,b})$ is the bare molecular propagator. The Fermi
distribution function is $f(x)=1/(e^{\beta x}+1)$.

Thus we have
\begin{eqnarray}
R(k, iw_n)&=&\frac{1}{\beta}\frac{f(\epsilon_{k, 3})}{iw_n+\epsilon_{k, 3}-\epsilon_{k,2}-
\Gamma(k,iw_n)},
\end{eqnarray}
with $\Gamma(k,iw_n)=\lambda^2\Sigma(k,iw_n+\epsilon_{k,3})$. The RF spectrum is
recovered as
\begin{eqnarray}
I(w)&&\propto {\rm Im}\frac{1}{V}\sum_kR(k, iw_n\to w+\mu+i0)\\ \nonumber
&&={\rm Im}\frac{1}{\beta}\int d^3k\frac{f(\epsilon_{k, 3})}{w+\mu+\epsilon_{k, 3}-\epsilon_{k,2}-
\Gamma(k,w+\mu)},
\label{eq:finittemp}
\end{eqnarray}
where in the last expression we implicitly assumed that $w$ has a small positive imaginary part.

At zero temperature, Eq. (\ref{eq:finittemp}) reduces to our previous approximation in Eq. (\ref{eq:zerotemp}).
In terms of dimensionless variables we write
\begin{eqnarray}
I(\tilde{w})\propto {\rm Im}\int_0^\infty d\tilde{k}\frac{\tilde{k}^2\tilde{f}(\tilde{k}^2-\tilde{\mu})}{\tilde{w}
-\tilde{\lambda}^2\tilde{\Sigma}}
\end{eqnarray}
with
\begin{equation}
\tilde{\Sigma}=\int_0^\infty d\tilde{p}\tilde{f}(\tilde{p}^2-\tilde{\mu})\frac{\tilde{p}}{\tilde{k}}
ln\left(\frac{\tilde{w}-\tilde{\delta}+(\tilde{k}+\tilde{p})^2/2}{\tilde{w}-\tilde{\delta}+(\tilde{k}-\tilde{p})^2/2}\right),
\end{equation}
where $\tilde{w}=w/\epsilon_F$, $\tilde{k}=k/k_F$, $\tilde{\mu}=\mu/\epsilon_F$, 
$\tilde{\beta}=\beta\epsilon_F$, $\tilde{\lambda}=\lambda\sqrt{k_F^3}/(2\pi\epsilon_F)$,
$\tilde{\delta}=\delta/\epsilon_F$ and $\tilde{f}(x)=1/(e^{\tilde{\beta}x}+1)$, where $\epsilon_F$ is an
arbitrary scale (typically taken to be the chemical potential), and $k_F^2/2m=\epsilon_F$.

\subsection{Results}

In Fig. \ref{fig:theory} we show the evolution of  the RF spectrum as a function of detuning and frequency 
for temperatures $T=0$, $\mu$, and $2\mu$. The chemical potential is fixed as $\tilde{\mu}=1$
and the coupling strength $\tilde{\lambda}=1$.  Many of the features in Fig. \ref{fig:theory} were anticipated
by our sum rule calculation in section II. A. There are two peaks, which disperse in opposite
directions, with spectral weight continuously shifting from one to another. A new feature, particularly
apparent at higher temperatures, is the peaks become quite broad near resonance. There is also a marked
asymmetry, where the negative energy peak is broader. One can attribute this broadening to the energy
dependence of the scattering. 

\section{comparison with experiment}
Here we compare the spectra in Fig. \ref{fig:theory} with the experiments in Ref. \cite{Ohara}. Rather than
modeling the harmonic trap, we will treat the gas as homogeneous, using the mean density in the 
experiment of Ref. \cite{Ohara}. Since the initial state is effectively non-interacting, we can
extract the chemical potential at a given temperature by solving $n=\int d^3kf(\epsilon_k-\mu)$. Our 
results are shown in Fig. \ref{fig:experiment} (cf. Fig. 4 from Ref. \cite{Ohara}). We see that on the scale
of fields used in the experiment, the bimodal structure is not apparent, and it is reasonable to model the
spectra by a single peak. At higher temperatures near resonance the peak is quite broad. The qualitative
position of the peak tracks well with the observations in Ref. \cite{Ohara}, but deviates quantitatively. The
discrepancies are likely attributable to inhomogeneities and uncertainty in the density.

\section{summary}
In summary, we have studied the RF spectrum of fermions near a narrow resonance. We presented a
sum rule calculation, which shows how the spectrum evolves from a wide to narrow resonance.  Wide
resonances possess a divergence which is cut off by the effective range.  We found bimodal behavior near
resonance. This bimodality becomes less apparent at high temperature and can be masked by
inhomogeneous broadening. At temperatures of order the Fermi temperature, both peaks broaden near
resonance. The positive energy peak, however, is distinctly sharper.

The RF lineshape teaches us at least two lessons about the underlying physics. First, as pointed out by
Kohstall {\it et al.}, the sharp positive detuning peak is consistent with the presence of a long-lived
repulsive polaron on the BEC side of resonance \cite{Kohstall}. Second, as already emphasized the
extreme broadness of the higher temperature spectra reveals the energy dependence of the scattering amplitude.

\begin{acknowledgements}
We thank K. O'Hara for discussions of experimental details. 
This research is supported by the National Science Foundation (PHY-1068165), the National Key Basic
Research Program of China (Grant No. 2013CB922000) and the National Natural Science Foundation
of China (Grant No. 11074021). J. X. is also supported by China Scholarship Council.
\end{acknowledgements}


\begin{thebibliography}{30}

\bibitem{Ketterle} S. Inouye, M. R. Andrews, J. Stenger, H.-J. Miesner, D. M. Stamper-Kurn, and W.
Ketterle, Nature {\bf 392}, 151 (1998).

\bibitem{Bloch} I. Bloch, J. Dalibard, and W. Zwerger, Rev. Mod. Phys. {\bf 80}, 885 (2008).

\bibitem{Chin} C. Chin, R. Grimm, P. Julienne, and E. Tiesinga, Rev. Mod. Phys. {\bf 82}, 1225 (2010).

\bibitem{Strecker} K. E. Strecker, G. B. Partridge, and R. G. Hulet, Phys. Rev. Lett. {\bf 91}, 080406 (2003).

\bibitem{Schwenk} A. Schwenk, and C. J. Pethick, Phys. Rev. Lett. {\bf 95}, 160401 (2005).

\bibitem{Petrov} D. S. Petrov, Phys. Rev. Lett. {\bf 93}, 143201 (2004).

\bibitem{Ho} T.-L. Ho, X. Cui, and W. Li, Phys. Rev. Lett. {\bf 108}, 250401 (2012).

\bibitem{Ohara} E. L. Hazlett, Y. Zhang, R. W. Stites, and K. M. O'Hara, Phys. Rev. Lett. {\bf 108}, 045304  (2012).

\bibitem{Jensen} L. M. Jensen, H. M. Nilsen, and G. Watanabe, Phys. Rev. A {\bf 74}, 043608 (2006).

\bibitem{Liu} M. M. Forbes, E. Gubankova, W. V. Liu, and F. Wilczek, Phys. Rev. Lett. {\bf 94}, 017001
(2005).

\bibitem{Kohstall} C. Kohstall, M. Zaccanti, M. Jag, A. Trenkwalder, P. Massignan, G. M. Bruun, F. Schreck, and R. Grimm, Nature {\bf 485}, 615 (2012).

\bibitem{Mueller1} K. R. A. Hazzard and E. J. Mueller, Phys. Rev. A {\bf 81}, 033404 (2010).

\bibitem{Mueller2} S. Basu and E. J. Mueller, Phys. Rev. Lett. {\bf 101}, 060405 (2008).

\bibitem{Stewart} J. T. Stewart, J. P. Gaebler, T. E. Drake, and D. S. Jin, Phys. Rev. Lett. {\bf 104}, 235301 (2010).

\bibitem{He} Y. He, C. C. Chien, Q. Chen, and K. Levin, Phys. Rev. Lett. {\bf 102}, 020402 (2009).

\bibitem{Pieri1} P. Pieri, A. Perali, and G. C. Strinati, Nature Physics {\bf 5}, 736 (2009). 

\bibitem{Haussmann} R. Haussmann, M. Punk, and W. Zwerger, Phys. Rev. A {\bf 80}, 063612 (2009).

\bibitem{Greiner} M. Greiner, C. A. Regal, and D. S. Jin, Phys. Rev. Lett. {\bf 94}, 070403 (2005).

\bibitem{Chin2} C. Chin, M. Bartenstein, A. Altmeyer, S. Riedl, S. Jochim, J. H. Denschlag, and R. Grimm, Science {\bf 305}, 1128 (2004).

\bibitem{Schunck} C. H. Schunck, Y. Shin, A. Schirotzek, M. W. Zwierlein, and W. Ketterle, Science {\bf 316}, 867 (2007).

\bibitem{Cheuk} L. W. Cheuk, A. T. Sommer, Z. Hadzibabic, T. Yefsah, W. S. Bakr, and M. W. Zwierlein, Phys. Rev. Lett. {\bf 109}, 095302 (2012).

\bibitem{Pieri2} P. Pieri, A. Perali, G. C. Strinati, S. Riedl, M. J. Wright, A. Altmeyer, C. Kohstall, E. R. Sanchez Guajardo, J. Hecker Denschlag, and R. Grimm, Phys. Rev. A {\bf 84}, 011608(R) (2011).

\bibitem{McMillan} W. L. McMillan and J. M. Rowell, Phys. Rev. Lett. {\bf 14}, 108 (1965).

\bibitem{Mueller} E. J. Mueller, Phys. Rev. A {\bf 78}, 045601 (2008).

\bibitem{Timmermans} E. Timmermans, P. Tommasini, M. Hussein, and A. Kerman, Phys. Rep. {\bf 315},
199 (1999).


\bibitem{Schneider} W. Schneider, V. B. Shenoy, and M. Randeria, arXiv: 0903.3006 (2009).

\end{thebibliography}
\end{document}